\documentstyle[twocolumn,aps,epsf,axodraw]{revtex}
\begin{document}
\title{Thermal effective potential of the linear sigma model}
\author{Nicholas Petropoulos}
\address{Theoretical Physics Group, Department of Physics and Astronomy\\
University of Manchester, Manchester, M13 9PL, UK\\}
\date{\today}
\maketitle
\begin{abstract}
We have attempted an approach to the chiral phase transition of QCD
using the linear sigma model as an effective theory. In order to get 
some insight  
into how the phase transition could proceed, we have 
calculated the finite temperature effective
potential of this model in the Hartree and large $N$ 
approximations using the 
Cornwall-Jackiw-Tomboulis formalism 
of composite operators.
\end{abstract}

\section{Introduction}
It is well known the chiral symmetry of QCD is broken due
to the small quark masses. A large amount of the current research
activity is devoted to  the chiral phase transition of QCD
by using effective models or lattice techniques. An account on  the physics
of thermal QCD and further references is given in    
\cite{smilga}. In studies of symmetry 
restoring phase transitions the 
finite temperature effective potential is an important 
and popular  theoretical tool \cite{doja,linderep}. The  
finite temperature effective potential 
$V(\phi,T)$  is defined through an effective 
action $\Gamma (\phi)$  which is the 
generating functional  of the one particle irreducible 
graphs and it has the meaning of 
the free energy density of the system under consideration.

A  generalised version is the 
effective potential $V(\phi,G)$ for
composite operators introduced by Cornwall, Jackiw and 
Tomboulis (CJT) \cite{cjt} and originally was written at zero
temperature but it has been extended to finite temperature 
by Amelino-Camelia and Pi \cite{gacpi}. In this case
the effective action $\Gamma(\phi,G)$ is the generating 
functional of the two particle irreducible 
vacuum graphs and the effective potential depends  
on $\phi(x)$ -- a possible expectation value of 
a quantum field  $\Phi(x)$ -- and on $G(x,y)$ -- a
possible expectation value of the time ordered product of the field 
operator $T\Phi(x)\Phi(y)$. 

Physical solutions demand that the effective potential  should satisfy
the stationarity requirments
\begin{equation}
\frac{d V(\phi,G)}{d\phi}=0\,\,\,
{\rm and}\,\,\,
\frac{dV(\phi,G)}{d G}=0~.
\end{equation}
\noindent  
The conventional effective potential results 
as $V(\phi)= V(\phi;G_{0})$ at the solution 
$G(\phi)=G_{0}(\phi)$ of the second equation.

There is an advantage in using the CJT method to calculate the 
effective potential in  Hartree approximation. As it was shown in
a recent investigation of the $\lambda\phi^{4}$ theory 
\cite{gacpi}, we need to evaluate 
only one type of graphs -- those in fig. 1a -- using 
an ansatz for a ``dressed propagator'', instead of summing  
the infinite class of ``daisy'' and 
``super daisy'' graphs -- figs. 1b, 1c  
respectively -- using  the usual 
tree level propagators.
\begin{figure}
\begin{center}
\begin{picture}(200,90)(105,0)
\BCirc(105,40){10}
\BCirc(105,60){10}
\Text(105,10)[c]{(a)}
\Text(180,10)[c]{(b)}
\Text(260,10)[c]{(c)}
\BCirc(180,50){20}
\CArc(197.7,67.7)(5,0,360)
\CArc(162.3,67.7)(5,0,360)
\CArc(197.7,32.3)(5,0,360)
\CArc(162.3,32.3)(5,0,360)
\BCirc(180,75){5}
\BCirc(180,25){5}
\BCirc(205,50){5}
\BCirc(155,50){5}
\CArc(197.7,67.7)(5,0,360)
\CArc(197.7,67.7)(5,0,360)
\CArc(197.7,67.7)(5,0,360)
\BCirc(255,40){20}
\BCirc(285,40){10}
\BCirc(300,40){5}
\BCirc(307.5,40){2.5}
\BCirc(285,55){5}
\BCirc(285,62.5){2.5}
\BCirc(255,70){10}
\BCirc(255,85){5}
\BCirc(255,92.5){2.5}
\BCirc(270,70){5}
\BCirc(240,70){5}
\end{picture}
\end{center}
\caption{The double bubble and examples of 
daisy and  superdaisy diagrams.}
\end{figure}
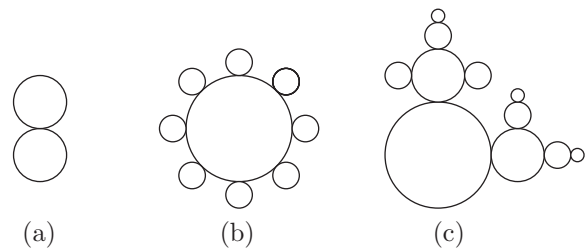

As in many other cases in physics, in order to deal with 
the chiral phase transition of QCD we can use effective models to describe
the physical situation. The linear sigma model \cite{Gell-Mann} 
is such a theory 
since it exhibits the correct chiral 
properties and  has attracted much attention recently, especially 
in studies of Disoriented Chiral 
Condensates \cite{bjorken,mikeaba,randrup-linear}.
The generalised version 
of the linear sigma model -- without fermions -- is called the $O(N)$
or vector model and is based on a set of 
$N$ real scalar fields. The $O(N)$ model Lagrangian 
can be written as
\begin{equation}
\label{nmodel}
{\cal L}=\frac{1}{2}(\partial_{\mu}{\bf \Phi})^{2}-
\frac{1}{2}m^{2}{\bf \Phi}^{2}
-\frac{1}{6N}\lambda {\bf \Phi}^{4}-\varepsilon\sigma~,
\label{lagrangianN}
\end{equation}
\noindent
and in order our notation to be consistent with applications on pion 
phenomenology we can identify the $\Phi_{1}$ with the $\sigma$
field and the remaining $N-1$ components as the pion 
fields, so ${\bf \Phi}=(\sigma, \pi_{1},\ldots,\pi_{N-1})$. The last 
term $-\varepsilon \sigma$
has been introduced in order to generate the  observed masses
of the pions.

The contact with phenomenology is obtained by considering 
the case $N=4$. Then the model 
consists of four scalar fields, the  
sigma $\sigma$ field  and 
the usual three pion fields 
$\pi^{0}$, $\pi^{\pm}$ 
which form a  four vector 
$(\sigma,\pi_{i})\;\;i=1,2,3$. The $\sigma$ field 
can be used to represent
the quark condensate, the order parameter for
the chiral phase transition,  since they exhibit the same behaviour
under chiral transformations \cite{mike}. The pions 
are very light particles and can
be considered  approximately as massless Goldstone 
bosons. In order to be consistent with the  observed pion  mass of
$m_{\pi}\approx 138\,\,\rm MeV$ and the usually 
adopted sigma mass of $m_{\sigma}\approx 600\,\,\rm MeV$
we choose $\varepsilon=f_{\pi}m_{\pi}^{2}$, where 
$f_{\pi}=93\,\,\rm MeV$ is the pion decay 
constant. The coupling constant $\lambda$  and the negative mass 
parameter $m^{2}$ of the  model are defined as
$\lambda=3(m_{\sigma}^{2}-m_{\pi}^{2})/f_{\pi}^{2}$ and
$-m^{2}=(m_{\sigma}^{2}-3 m_{\pi}^{2})/2 > 0$.

\section{The Hartree approximation}
We examine the $O(4)$ model first by setting  $N=4$ in 
the Lagrangian (\ref{lagrangianN}). In the chiral limit when 
the pions are considered as massless
the last term  is ignored. The four types of bubble diagrams relevant 
for Hartree
approximation are given in fig. 2.
\begin{figure}
\label{bubbles}
\begin{center}
\begin{picture}(240,60)(20,20)
\BCirc(85,55){10}
\BCirc(85,75){10}
\Text(85,40)[c]{$\sigma$}
\Text(85,90)[c]{$\sigma$}
\Text(85,30)[c]{$3$}
\BCirc(125,55){10}
\Text(125,40)[c]{$\pi_{i}$}
\BCirc(125,75){10}
\Text(125,90)[c]{$\pi_{i}$}
\Text(125,30)[c]{$9$}
\BCirc(165,55){10}
\Text(165,40)[c]{$\pi_{i}$}
\BCirc(165,75){10}
\Text(165,90)[c]{$\pi_{j\neq i}$}
\Text(165,30)[c]{$6$}
\BCirc(205,55){10}
\Text(205,40)[c]{$\sigma$}
\BCirc(205,75){10}
\Text(205,90)[c]{$\pi_{i}$}
\Text(205,30)[c]{$6$}
\end{picture}
\caption{Graphs which contribute to
the effective potential for the O(4) linear 
sigma model in the Hatree approximation. The numbers show
the weight of each type of bubble in the expression for
the effective potential.}   
\end{center}
\end{figure}
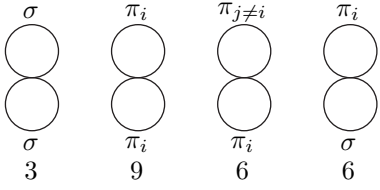

Following the procedure described in \cite{nikos}
the thermal effective potential reads as
\begin{eqnarray}
V(\phi,M) &=& \frac{1}{2}m^{2}\phi^{2}+\frac{1}{24}\lambda\phi^{4}
          +{1 \over 2}\int_{\beta}\ln G^{-1}_{\sigma}(k)\nonumber\\
&+&{3 \over 2}\int_{\beta}\ln G^{-1}_{\pi}(k)
+{1 \over 2}\int_{\beta}
             [{\cal D}_{\sigma}^{-1}(k)G_{\sigma}(k)-1]\nonumber\\
          &+&{3 \over 2}\int_{\beta}
              [{\cal D}_{\pi}^{-1}(k)G_{\pi}(k)-3]\nonumber\\
          &+&3\frac{\lambda}{24}\left [
\int_{\beta}G_{\sigma}(k)\right ]^{2}
+15\frac{\lambda}{24}\left [
\int_{\beta}G_{\pi}(k)\right ]^{2}\nonumber\\
&+&6\frac{\lambda}{24}\left [\int_{\beta}G_{\sigma}(k)\right ]
\left [\int_{\beta}G_{\pi}(k)\right ]~.
\end{eqnarray}
In our calculations we use the imaginary time formalism 
and for simplicity we use the following notation
\begin{equation}
\int\frac{d^{4}k}{(2\pi)^{4}}\longrightarrow
 \frac{1}{\beta}\sum_{n}^{\infty}\int\frac{d^{3}\rm k}{(2\pi)^{3}}
\equiv\int_{\beta}.
\end{equation}
The tree propagators appearing above are given by
\begin{eqnarray}
& &{\cal D}^{-1}_{\sigma}(\phi;k) = k^2 +m^2 + {1\over 2}\lambda\phi^2 
\nonumber\\
& &{\cal D}^{-1}_{\pi}(\phi;k) = k^2 +m^2 + {1\over 6}\lambda\phi^2 ~,  
\end{eqnarray}
and as in \cite{gacpi} we adopt the following form
for the {\it dressed propagator}
\begin{equation}
G_{\sigma/\pi}^{-1}(\phi;k)=k^{2}+M^{2}_{\sigma/\pi} 
\end{equation}
introducing a thermal effective mass $M_{\sigma/\pi}$.
Minimizing the potential with respect to $G_{\sigma/\pi}$
we obtain a system of
gap equations for the thermal effective masses, which after performing
the Matsubara frequency sums appears as
\begin{eqnarray}
&&M_{\sigma}^{2}=m^{2}+\frac{1}{2}\lambda\phi^{2}+
\frac{\lambda}{2}F_{\beta}(M_{\sigma})+\frac{\lambda}{2}F_{\beta}(M_{\pi})
\nonumber\\
&&M_{\pi}^{2}=m^{2}+\frac{1}{6}\lambda\phi^{2}
+\frac{\lambda}{6}F_{\beta}(M_{\sigma})+\frac{5\lambda}{6}F_{\beta}(M_{\pi})~.
\label{masses4}
\end{eqnarray}
and at the level of our approximation we 
keep only the finite temperature part of the resulting integrals, so
$F_{\beta}(M_{\pi/\sigma})$ is given by
\begin{equation}
F_{\beta}(M)=\frac{T^{2}}{2\pi^{2}}\int_{0}^{\infty}\frac{x^2 dx}
{a^{1/2}}
\frac{1}{\exp(a^{1/2})-1}~,
\label{FM}
\end{equation}
with $a=x^{2}+y^{2}$ and  $y=M_{\pi/\sigma}/T$.
Minimizing with respect to the  order parameter $\phi$
we find one more equation
\begin{equation}
0=m^{2}+\frac{1}{6}\lambda\phi^{2}+\frac{\lambda}{2}F_{\beta}(M_{\sigma})
+\frac{\lambda}{2}F_{\beta}(M_{\pi})~.
\end{equation}
This system is solved numerically and the solution is given in fig. 3
\begin{figure}
\begin{center}
\mbox{
\epsfxsize=6cm
\epsfbox{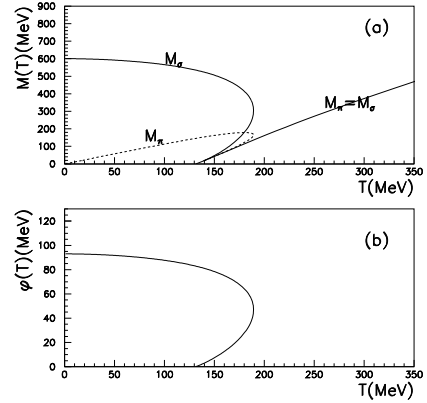}
}
\end{center}
\caption{(a) The
sigma and pion effective masses  as functions
 of temperature. (b) Order parameter 
as a function of temperature.}
\end{figure}

The finite temperature effective potential $V(\phi,T)$ written in compact
form is given by
\begin{eqnarray*}
V(\phi,M)&=&\frac{1}{2}m^{2}\phi^{2}+\frac{1}{24}\lambda\phi^{4}
+ Q_{\beta}(M_{\sigma})\nonumber\\
&&+ 3Q_{\beta}(M_{\pi})
-\frac{\lambda}{8}[F_{\beta}(M_{\sigma})]^{2}\nonumber\\
&&-\frac{5\lambda}{8}[F_{\beta}(M_{\pi})]^{2}
-\frac{\lambda}{4}F_{\beta}(M_{\sigma})F_{\beta}(M_{\pi})~,
\end{eqnarray*}
where $F_{\beta}(M)$ 
is given by the equation (\ref{FM}) and 
\begin{equation}
Q_{\beta}(M)=\frac{T^4}{2\pi^2}\int_{0}^{\infty}dx x^2 
\ln\left [1-\exp(-a^{1/2})\right ]
\label{QM}
\end{equation}
with $a=x^2+y^2$ and $y=M_{\pi/\sigma}/T$. The evolution of the 
potential with temperature
is given in fig. 4 and we 
easily observe indication of a first order phase
transition.
\begin{figure}
\label{critical}
\begin{center}
\mbox{
\epsfxsize=6cm
\epsfbox{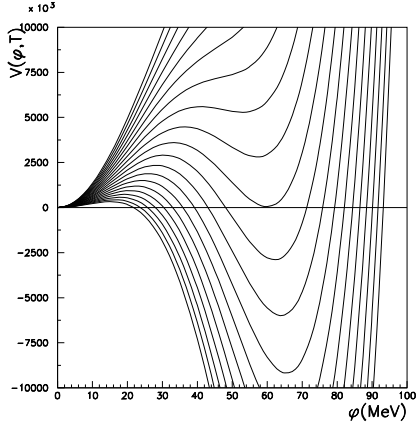}
}
\end{center}
\caption{Evolution of the effective 
potential $V(\phi,T)$ as 
a function of the order parameter $\phi$ for several temperatures
in steps of $2\;{\rm MeV}$. The two minima appear 
as degenerate at $T_{c}\approx 182\;{\rm MeV}$. The second minimum
at  $\phi \neq 0$ dissapears at a 
temperature $T_{c2}\approx 187\;{\rm MeV}.$}
\end{figure}


When the chiral symmetry is broken minimizing the thermal effective
potential with respect to dressed propagators we find the
same set of gap equations (\ref{masses4}) for the 
effective masses as before, but minimizing
with respect to the order parameter we obtain the following equation
\begin{equation}
[m^{2}+\frac{1}{6}\lambda\phi^{2}+\frac{\lambda}{2}F_{\beta}(M_{\sigma})
+\frac{\lambda}{2}F_{\beta}(M_{\pi})]\phi-\varepsilon =0 ~. 
\label{phi4}
\end{equation}
We have solved the system of equations (\ref{masses4}) and (\ref{phi4})
numerically and the solution is given in fig. 5.
\begin{figure}
\begin{center}
\mbox{
\epsfxsize=6cm
\epsfbox{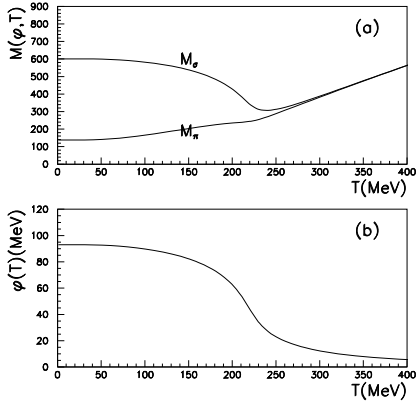}
}
\end{center}
\caption{(a) Solution of the system of gap 
equations in the case 
when $\epsilon \neq 0$. At low temp the pions appear with
the observed masses. (b) Evolution of the order parameter as a function 
of temperature.}
\end{figure}
There is no longer
any phase transition. We rather observe a crossover
phenomenon from the low temperature phase to the high temperature
phase.   
\section{Large $N$ approximation}

When we consider $N-1$ pion fields the Lagrangian of the model
is given in equation (\ref{lagrangianN}) and in the chiral limit
we ignore the last term. Then the thermal 
effective potential looks like \cite{nikos,gac97}
\begin{eqnarray}
V(\phi,M) &=&\frac{1}{2}m^{2}\phi^{2}+\frac{1}{6N}\lambda\phi^{4}
+{N-1\over 2}\int_{\beta}\ln G^{-1}_{\pi}(k)\nonumber\\
&+&{1 \over 2}\int_{\beta}\ln  G^{-1}_{\sigma}(k)
+{1 \over 2}\int_{\beta}[{\cal D}_{\sigma}^{-1}(k)G_{\sigma}(k)-1]
\nonumber\\
&+&{N-1 \over 2}\int_{\beta}
              [{\cal D}_{\pi}^{-1}(k)G_{\pi}(k)-(N-1)]\nonumber\\
&+&3\frac{\lambda}{24}\left [\int_{\beta}G_{\sigma}(k)\right ]^{2}
+\frac{\lambda(N^{2}-1)}{6N}\left [
\int_{\beta} G_{\pi}(k)\right ]^{2}\nonumber\\
&+&\frac{\lambda(N-1)}{6N}\left [\int_{\beta} G_{\sigma}(k)\right ]
\left [\int_{\beta}G_{\pi}(k)\right ]~.
\end{eqnarray}
Minimizing the potential with respect to the dressed propagators we find 
a set of  nonlinear gap equations 
which in the large $N$ approximation reduces to
\begin{eqnarray}
&&M_{\sigma}^{2}=m^{2}+\frac{1}{2}\lambda\phi^{2}
+\frac{2\lambda}{3}F_{\beta}(M_{\pi})
\nonumber\\
&&M_{\pi}^{2}=m^{2}+\frac{1}{6}\lambda\phi^{2}
+\frac{2\lambda}{3}F_{\beta}(M_{\pi})~.
\label{massesN}
\end{eqnarray}
Minimizing the potential with respect to $\phi$ we obtain
\begin{equation}
{dV(\phi,M)\over d\phi}=\phi M^{2}_{\pi}=0
\end{equation}
which means that in this approximation the 
Goldstone theorem is satisfied. The solution of the system is given in fig. 6.
\begin{figure}
\begin{center}
\mbox{
\epsfxsize=6cm
\epsfbox{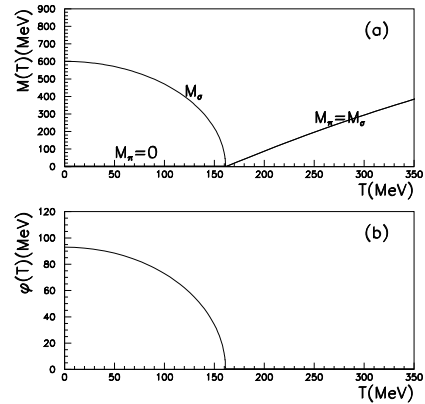}
}
\end{center}
\caption{(a) Solution of the system of 
gap equations in the large $N$
approximation in the  chiral limit. At low temperatures
the pions appear as massless.(b) Evolution of the order parameter with
temperature in the large $N$ approximation.}
\end{figure}
In the high temperature phase when $\phi=0$, the two equations become
degenerate and can be used to calculate the critical temperature 
$T_{c}=\sqrt{3}f_{\pi}\approx 161 \rm MeV$. 

The presence of the symmetry breaking term in the 
Lagrangian (\ref{lagrangianN}) results -- after minimizing the
potential -- the same set of equations (\ref{massesN}) as before, but minimizing 
with respect to $\phi$ we obtain the equation
\begin{equation}
[m^{2}+\frac{1}{6}\lambda\phi^{2}
+\frac{2\lambda}{3}F_{\beta}(M_{\pi})]\phi-\varepsilon=0~.
\label{phiN}
\end{equation}

We have solved the system of equations (\ref{massesN}) and (\ref{phiN})
numerically and the solution is given in fig 7.
\begin{figure}
\begin{center}
\mbox{
\epsfxsize=6cm
\epsfbox{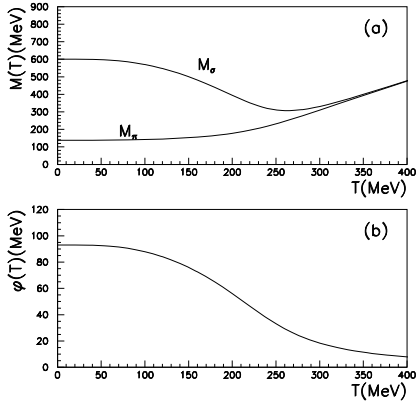}
}
\end{center}
\caption{(a) Solution of the system of 
gap equations in the large $N$
approximation in the case of broken chiral symmetry
$\varepsilon \neq 0$. At low temperatures the pions appear with
the observed masses. (b) Evolution of the order parameter with
temperature.}
\end{figure}
As is the Hartree case 
we do not observe a phase transition. We encounter the 
crossover 
phenomenon again; the
difference being now the smoother behaviour of the order
parameter. 
\section{Conclusions}
The calculation of the thermal effective
potential of the linear sigma model in the Hartree and large $N$ 
approximations using the CJT  formalism 
of composite operators proved to be very handy since we 
actually need 
to calculate only one type of diagram. 
In both cases we 
have solved numerically the system
of the resulting  gap equations and found  the evolution with temperature of 
the thermal effective  masses. In the 
chiral limit, when the pion mass
is ignored, the Hartree 
approximation  predicts a first order phase 
transition, but in contrast to that, in the large $N$
approximation we find a second order phase transition. This
last observation seems to be in agreement with
different approaches to the chiral phase transition based 
on the argument that the linear sigma model belongs
to the same  universality class as other models which
are known  to exhibit second order phase 
transitions \cite{rajagopal-wilczek}. However in the large
$N$ approximation the sigma contribution is ignored and
this of course introduces errors when we
calculate  the critical temperature. In the $N=4$ case
which is closer to phenomenology we could probably obtain 
a better approximation in calculating the transition
temperature  if we had
considered all the interactions described by the
Lagrangian of the model. We are planning
some investigations in order  to include these effects 
in the calculation of the effective potential.	 

When we introduce  a symmetry breaking 
term, $\varepsilon \sigma$, which generates 
the observed pion 
masses in the model,  both in the Hartree and 
large $N$ approximationds we found that there is no longer any phase
transition. We rather  observe a 
crossover phenomenon which
confirms other results reported recently using the linear sigma
model \cite{chiku}. In this analysis there is also an
indication of a first order phase transition in the chiral limit.
First order phase transition is also reported in \cite{randrup-linear,matsui}

I would like to express my graditude to Mike Birse
for his general help throughout
this work and the theory group in Manchester
for providing me financial support to attend 
this workshop. I would like also to thank the organisers, Prof Peter
Landshoff for a helpfull discussion and suggestions and J. Knoll, 
A. Smilga  and M. Tytgat for useful comments. Fruitful discussions with Antonio
Filippi and Massimo Blassone are also acknowledged.

\end{document}